\documentclass{sig-alternate}

\usepackage{amsfonts}
\usepackage{amsmath}
\usepackage{amssymb}
\usepackage{array}
\usepackage{arydshln}
\usepackage[english]{babel}
\usepackage[utf8]{inputenc}  \usepackage{ifthen}
\usepackage{csquotes}
\usepackage{courier}
\usepackage{enumitem}
\usepackage{fancyvrb}
\usepackage{float}
\usepackage{graphicx}
\usepackage[scaled]{helvet}
\usepackage{listings}
\usepackage{mathtools}
\usepackage{subfig}
\usepackage{tabularx}
\usepackage{textcomp}
\usepackage{tikz}
\usepackage{url}
\usepackage{wrapfig}
\usepackage{relsize}
\usepackage{xcolor}
\usepackage{xspace}
\usepackage{pgfplots}
\newcommand{\SBDEPARTMENT}{Computer Science Department}
\newcommand{\SBUNIVERSITY}{Saarland University}
\newcommand{\SAARBRUECKEN}{Saarbr{\"u}cken, Germany}

\newcommand{\TITLE}{\title{Polly's Polyhedral Scheduling in the Presence of Reductions}}
\newcommand{\SUBTITLE}{\subtitle{}}

\newcommand{\AUTHORS}{
  \numberofauthors{2}

  \author{
            \alignauthor
    Johannes Doerfert, Kevin Streit, and Sebastian Hack\\
    \affaddr{\SBDEPARTMENT}\\
    \affaddr{\SBUNIVERSITY}\\
    \affaddr{\SAARBRUECKEN}\\
    \email{<lastname>@cs.uni-saarland.de}
            \alignauthor
    Zino Benaissa\\
    \affaddr{Qualcomm Innovation Center}\\
    \affaddr{San Diego}\\
    \affaddr{California, USA}\\
    \email{zinob@quicinc.com}
  }
}

\toappear{
  \hrule \vspace{5pt}
  IMPACT 2015\\
  Fifth International Workshop on Polyhedral Compilation Techniques\\
  Jan 19, 2015, Amsterdam, The Netherlands\\
  In conjunction with HiPEAC 2015.\\[10pt]
  \url{http://impact.gforge.inria.fr/impact2015}\\
}

 \conferenceinfo{IMPACT}{2015 Amsterdam, Netherlands}

\newcommand{\BIBLIOGRAPHYSTYLE}{\bibliographystyle{abbrv}}
\newcommand{\BIBLIOGRAPHYFILE}{\bibliography{paper}}

\newcommand{\PREABSTRACT}{}
\newcommand{\POSTABSTRACT}{

    \category{D 3.4}{Programming languages}{Processors}[Compilers, Optimization]

  \terms{Algorithms; Performance}

  \keywords{Compiler Optimization; Affine Scheduling;
  Reductions}
}

\makeatletter
\def\PY@reset{\let\PY@it=\relax \let\PY@bf=\relax%
    \let\PY@ul=\relax \let\PY@tc=\relax%
    \let\PY@bc=\relax \let\PY@ff=\relax}
\def\PY@tok#1{\csname PY@tok@#1\endcsname}
\def\PY@toks#1+{\ifx\relax#1\empty\else%
    \PY@tok{#1}\expandafter\PY@toks\fi}
\def\PY@do#1{\PY@bc{\PY@tc{\PY@ul{%
    \PY@it{\PY@bf{\PY@ff{#1}}}}}}}
\def\PY#1#2{\PY@reset\PY@toks#1+\relax+\PY@do{#2}}

\expandafter\def\csname PY@tok@gd\endcsname{\def\PY@tc##1{\textcolor[rgb]{0.63,0.00,0.00}{##1}}}
\expandafter\def\csname PY@tok@gu\endcsname{\let\PY@bf=\textbf\def\PY@tc##1{\textcolor[rgb]{0.50,0.00,0.50}{##1}}}
\expandafter\def\csname PY@tok@gt\endcsname{\def\PY@tc##1{\textcolor[rgb]{0.00,0.27,0.87}{##1}}}
\expandafter\def\csname PY@tok@gs\endcsname{\let\PY@bf=\textbf}
\expandafter\def\csname PY@tok@gr\endcsname{\def\PY@tc##1{\textcolor[rgb]{1.00,0.00,0.00}{##1}}}
\expandafter\def\csname PY@tok@cm\endcsname{\let\PY@it=\textit\def\PY@tc##1{\textcolor[rgb]{0.25,0.50,0.50}{##1}}}
\expandafter\def\csname PY@tok@vg\endcsname{\def\PY@tc##1{\textcolor[rgb]{0.10,0.09,0.49}{##1}}}
\expandafter\def\csname PY@tok@m\endcsname{\def\PY@tc##1{\textcolor[rgb]{0.40,0.40,0.40}{##1}}}
\expandafter\def\csname PY@tok@mh\endcsname{\def\PY@tc##1{\textcolor[rgb]{0.40,0.40,0.40}{##1}}}
\expandafter\def\csname PY@tok@go\endcsname{\def\PY@tc##1{\textcolor[rgb]{0.53,0.53,0.53}{##1}}}
\expandafter\def\csname PY@tok@ge\endcsname{\let\PY@it=\textit}
\expandafter\def\csname PY@tok@vc\endcsname{\def\PY@tc##1{\textcolor[rgb]{0.10,0.09,0.49}{##1}}}
\expandafter\def\csname PY@tok@il\endcsname{\def\PY@tc##1{\textcolor[rgb]{0.40,0.40,0.40}{##1}}}
\expandafter\def\csname PY@tok@cs\endcsname{\let\PY@it=\textit\def\PY@tc##1{\textcolor[rgb]{0.25,0.50,0.50}{##1}}}
\expandafter\def\csname PY@tok@cp\endcsname{\def\PY@tc##1{\textcolor[rgb]{0.74,0.48,0.00}{##1}}}
\expandafter\def\csname PY@tok@gi\endcsname{\def\PY@tc##1{\textcolor[rgb]{0.00,0.63,0.00}{##1}}}
\expandafter\def\csname PY@tok@gh\endcsname{\let\PY@bf=\textbf\def\PY@tc##1{\textcolor[rgb]{0.00,0.00,0.50}{##1}}}
\expandafter\def\csname PY@tok@ni\endcsname{\let\PY@bf=\textbf\def\PY@tc##1{\textcolor[rgb]{0.60,0.60,0.60}{##1}}}
\expandafter\def\csname PY@tok@nl\endcsname{\def\PY@tc##1{\textcolor[rgb]{0.63,0.63,0.00}{##1}}}
\expandafter\def\csname PY@tok@nn\endcsname{\let\PY@bf=\textbf\def\PY@tc##1{\textcolor[rgb]{0.00,0.00,1.00}{##1}}}
\expandafter\def\csname PY@tok@no\endcsname{\def\PY@tc##1{\textcolor[rgb]{0.53,0.00,0.00}{##1}}}
\expandafter\def\csname PY@tok@na\endcsname{\def\PY@tc##1{\textcolor[rgb]{0.49,0.56,0.16}{##1}}}
\expandafter\def\csname PY@tok@nb\endcsname{\def\PY@tc##1{\textcolor[rgb]{0.00,0.50,0.00}{##1}}}
\expandafter\def\csname PY@tok@nc\endcsname{\let\PY@bf=\textbf\def\PY@tc##1{\textcolor[rgb]{0.00,0.00,1.00}{##1}}}
\expandafter\def\csname PY@tok@nd\endcsname{\def\PY@tc##1{\textcolor[rgb]{0.67,0.13,1.00}{##1}}}
\expandafter\def\csname PY@tok@ne\endcsname{\let\PY@bf=\textbf\def\PY@tc##1{\textcolor[rgb]{0.82,0.25,0.23}{##1}}}
\expandafter\def\csname PY@tok@nf\endcsname{\def\PY@tc##1{\textcolor[rgb]{0.00,0.00,1.00}{##1}}}
\expandafter\def\csname PY@tok@si\endcsname{\let\PY@bf=\textbf\def\PY@tc##1{\textcolor[rgb]{0.73,0.40,0.53}{##1}}}
\expandafter\def\csname PY@tok@s2\endcsname{\def\PY@tc##1{\textcolor[rgb]{0.73,0.13,0.13}{##1}}}
\expandafter\def\csname PY@tok@vi\endcsname{\def\PY@tc##1{\textcolor[rgb]{0.10,0.09,0.49}{##1}}}
\expandafter\def\csname PY@tok@nt\endcsname{\let\PY@bf=\textbf\def\PY@tc##1{\textcolor[rgb]{0.00,0.50,0.00}{##1}}}
\expandafter\def\csname PY@tok@nv\endcsname{\def\PY@tc##1{\textcolor[rgb]{0.10,0.09,0.49}{##1}}}
\expandafter\def\csname PY@tok@s1\endcsname{\def\PY@tc##1{\textcolor[rgb]{0.73,0.13,0.13}{##1}}}
\expandafter\def\csname PY@tok@sh\endcsname{\def\PY@tc##1{\textcolor[rgb]{0.73,0.13,0.13}{##1}}}
\expandafter\def\csname PY@tok@sc\endcsname{\def\PY@tc##1{\textcolor[rgb]{0.73,0.13,0.13}{##1}}}
\expandafter\def\csname PY@tok@sx\endcsname{\def\PY@tc##1{\textcolor[rgb]{0.00,0.50,0.00}{##1}}}
\expandafter\def\csname PY@tok@bp\endcsname{\def\PY@tc##1{\textcolor[rgb]{0.00,0.50,0.00}{##1}}}
\expandafter\def\csname PY@tok@c1\endcsname{\let\PY@it=\textit\def\PY@tc##1{\textcolor[rgb]{0.25,0.50,0.50}{##1}}}
\expandafter\def\csname PY@tok@kc\endcsname{\let\PY@bf=\textbf\def\PY@tc##1{\textcolor[rgb]{0.00,0.50,0.00}{##1}}}
\expandafter\def\csname PY@tok@c\endcsname{\let\PY@it=\textit\def\PY@tc##1{\textcolor[rgb]{0.25,0.50,0.50}{##1}}}
\expandafter\def\csname PY@tok@mf\endcsname{\def\PY@tc##1{\textcolor[rgb]{0.40,0.40,0.40}{##1}}}
\expandafter\def\csname PY@tok@err\endcsname{\def\PY@bc##1{\setlength{\fboxsep}{0pt}\fcolorbox[rgb]{1.00,0.00,0.00}{1,1,1}{\strut ##1}}}
\expandafter\def\csname PY@tok@kd\endcsname{\let\PY@bf=\textbf\def\PY@tc##1{\textcolor[rgb]{0.00,0.50,0.00}{##1}}}
\expandafter\def\csname PY@tok@ss\endcsname{\def\PY@tc##1{\textcolor[rgb]{0.10,0.09,0.49}{##1}}}
\expandafter\def\csname PY@tok@sr\endcsname{\def\PY@tc##1{\textcolor[rgb]{0.73,0.40,0.53}{##1}}}
\expandafter\def\csname PY@tok@mo\endcsname{\def\PY@tc##1{\textcolor[rgb]{0.40,0.40,0.40}{##1}}}
\expandafter\def\csname PY@tok@kn\endcsname{\let\PY@bf=\textbf\def\PY@tc##1{\textcolor[rgb]{0.00,0.50,0.00}{##1}}}
\expandafter\def\csname PY@tok@mi\endcsname{\def\PY@tc##1{\textcolor[rgb]{0.40,0.40,0.40}{##1}}}
\expandafter\def\csname PY@tok@gp\endcsname{\let\PY@bf=\textbf\def\PY@tc##1{\textcolor[rgb]{0.00,0.00,0.50}{##1}}}
\expandafter\def\csname PY@tok@o\endcsname{\def\PY@tc##1{\textcolor[rgb]{0.40,0.40,0.40}{##1}}}
\expandafter\def\csname PY@tok@kr\endcsname{\let\PY@bf=\textbf\def\PY@tc##1{\textcolor[rgb]{0.00,0.50,0.00}{##1}}}
\expandafter\def\csname PY@tok@s\endcsname{\def\PY@tc##1{\textcolor[rgb]{0.73,0.13,0.13}{##1}}}
\expandafter\def\csname PY@tok@kp\endcsname{\def\PY@tc##1{\textcolor[rgb]{0.00,0.50,0.00}{##1}}}
\expandafter\def\csname PY@tok@w\endcsname{\def\PY@tc##1{\textcolor[rgb]{0.73,0.73,0.73}{##1}}}
\expandafter\def\csname PY@tok@kt\endcsname{\def\PY@tc##1{\textcolor[rgb]{0.69,0.00,0.25}{##1}}}
\expandafter\def\csname PY@tok@ow\endcsname{\let\PY@bf=\textbf\def\PY@tc##1{\textcolor[rgb]{0.67,0.13,1.00}{##1}}}
\expandafter\def\csname PY@tok@sb\endcsname{\def\PY@tc##1{\textcolor[rgb]{0.73,0.13,0.13}{##1}}}
\expandafter\def\csname PY@tok@k\endcsname{\let\PY@bf=\textbf\def\PY@tc##1{\textcolor[rgb]{0.00,0.50,0.00}{##1}}}
\expandafter\def\csname PY@tok@se\endcsname{\let\PY@bf=\textbf\def\PY@tc##1{\textcolor[rgb]{0.73,0.40,0.13}{##1}}}
\expandafter\def\csname PY@tok@sd\endcsname{\let\PY@it=\textit\def\PY@tc##1{\textcolor[rgb]{0.73,0.13,0.13}{##1}}}

\def\PYZus{\char`\_}
\def\PYZob{\char`\{}
\def\PYZcb{\char`\}}

\def\PYZlt{\char`\<}

\def\PYZhy{\char`\-}

\makeatother

 \usetikzlibrary{
  arrows,
  shapes.misc,
  shapes.arrows,
  chains,
  matrix,
  positioning,
  patterns,
  scopes,
  decorations.pathmorphing,
  shadows
}

\tikzset{
  nonterminal/.style={
        rectangle,
        minimum size=6mm,
        very thick,
    draw=red!50!black!50,                                                   top color=white,                  bottom color=red!50!black!20,         font=\itshape
  },
  terminal/.style={
        rounded rectangle,
    minimum size=6mm,
        very thick,draw=black!50,
    top color=white,bottom color=black!20,
    font=\ttfamily},
  skip loop/.style={to path={-- ++(0,#1) -| (\tikztotarget)}}
}

\newcommand{\TODOBGCOLOR}{yellow}
\newcommand{\TODOFGCOLOR}{red}
\newcommand{\TODO}[2]{
  \ifthenelse{\equal{#1}{JD}}{
    \renewcommand{\TODOBGCOLOR}{green}
    \renewcommand{\TODOFGCOLOR}{red}
  }{

  }
  \colorbox{\TODOBGCOLOR}{    \begin{minipage}{0.98\linewidth}      \textcolor{\TODOFGCOLOR}{\small\textbf{#1:} #2}    \end{minipage}  }}

\mathtoolsset{centercolon}

\definecolor{myred}{HTML}{D7191C}
\definecolor{myorgange}{HTML}{FDAE61}
\definecolor{myyellow}{HTML}{FFFFBF}
\definecolor{mygreen}{HTML}{ABDDA4}
\definecolor{myblue}{HTML}{2B83BA}

\newcommand{\llvm}{\emph{LLVM}\xspace}
\newcommand{\polly}{\emph{Polly}\xspace}

\newcommand{\pscop}{\emph{SCoP}\xspace}
\newcommand{\pscops}{\emph{SCoP}s\xspace}

\let\bar=\|
\newcommand{\mathid}[1]{\text{\rmfamily\textit{#1}}}
\def\|#1|{\mathid{#1}}

\newcommand{\latticeO}{\text{$\mathcal{R}_{op}$\xspace}}
\newcommand{\latticeL}{\text{$loads(S)$\xspace}}
\newcommand{\redLike}{\text{$R_{c}$\xspace}}

\newcommand{\redLikes}{\text{$\mathcal{R}_{c}$\xspace}}

\newcommand{\loadI}{{\tt load}\xspace}
\newcommand{\loadIl}{{\tt load\ {}}{\tt l}\xspace}

\newcommand{\storeI}{{\tt store}\xspace}
\newcommand{\storeIs}{{\tt store\ {}}{\tt s}\xspace}

\newcommand{\tf}{{\it \text{$t_S$}}\,{}}

\newcommand{\Insts}{{\tt Insts}\xspace}
\newcommand{\DepSet}{\(\mathcal{D}\)\xspace}
\newcommand{\DepSetRed}{\(\mathcal{D}_\rho\)\xspace}
\newcommand{\DepSetNonRed}{\(\mathcal{D}_\nu\)\xspace}
\newcommand{\DepSetPriv}{\(\mathcal{D}_\tau\)\xspace}
\newcommand{\DepSetMath}{\mathcal{D}}
\newcommand{\DepSetRedMath}{\mathcal{D}_\rho}
\newcommand{\DepSetNonRedMath}{\mathcal{D}_\nu}
\newcommand{\DepSetPrivMath}{\mathcal{D}_\tau}

\begin{document}

\TITLE
\SUBTITLE
\AUTHORS

\maketitle

\PREABSTRACT
\begin{abstract}

The polyhedral model provides a powerful mathematical abstraction to enable
effective optimization of loop nests with respect to a given optimization goal,
e.g., exploiting parallelism. Unexploited reduction properties are a frequent
reason for polyhedral optimizers to assume parallelism prohibiting dependences.
To our knowledge, no polyhedral loop optimizer available in any production
compiler provides support for reductions. In this paper, we show that leveraging
the parallelism of reductions can lead to a significant performance increase. We
give a precise, dependence based, definition of reductions and discuss ways to
extend polyhedral optimization to exploit the associativity and commutativity of
reduction computations. We have implemented a reduction-enabled scheduling
approach in the Polly polyhedral optimizer and evaluate it on the standard
Polybench 3.2 benchmark suite. We were able to detect and model all 52
arithmetic reductions and achieve speedups up to 2.21× on a quad core machine by
exploiting the multidimensional reduction in the BiCG benchmark.

\end{abstract}
 \POSTABSTRACT

\section{Introduction} \label{sec:Introduction}

Over the last four decades various
approaches~\cite{Jouvelot1986,Jouvelot1989,Blelloch1989,Pinter1991,Redon1993,Fisher1994,Pottenger1995,Rauchwerger1995,Suganuma1996,Yu2004,Gautam2006}
were proposed to tackle reductions: a computational idiom which prevents
parallelism due to loop carried data dependences. An often used definition for
reductions describes them as an associative and commutative computation which
reduces the dimensionality of a set of input data~\cite{Midkiff2012}. A simple
example is the array sum depicted in
Figure~\ref{fig:array_sum_sequential_code}. The input vector {\tt A} is reduced
to the scalar variable {\tt sum} using the associative and commutative operator
{\tt +}. In terms of data dependences, the loop has to be computed sequentially
because a read of the variable {\tt sum} in iteration~$i+1$ depends on the
value written in iteration~$i$. However, the associativity and commutativity of
the reduction operator can be exploited to reorder, parallelize or vectorize
such reductions.

While reordering the reduction iterations is always a valid transformation,
executing reductions in a parallel context requires additional ``fix up''.
Static transformations often use privatization as fix up technique as it
works well with both small and large parallel tasks.
The idea of privatization is to duplicate the shared memory locations for each
instance running in parallel. Thus, each parallel instance works on a private
copy of a shared memory location. Using the privatization scheme we can
vectorize the array sum example as shown in
Figure~\ref{fig:array_sum_vectorized_code}. For the shared variable {\tt sum}, a
temporary array {\tt tmp\_sum}, with as many elements as there are vector lanes,
is introduced. Now the computation for each vector lane uses one array element
to accumulate intermediate results unaffected by the computations of the other
lanes. As the reduction computation is now done in the temporary array instead
of the original reduction location we finally need to accumulate all
intermediate results into the original reduction location. This way,
users of the variable {\tt sum} will still see the overall sum of all array
elements, even though it was computed in partial sums first.

\begin{figure}[htbp]
  \centering
  \subfloat[Sequential array sum computation.]{
    \input{array_sum_sequential.tex}
    \label{fig:array_sum_sequential_code}
  }

  \subfloat[Vectorized array sum computation.]{
    \input{array_sum_vectorized.tex}
    \label{fig:array_sum_vectorized_code}
  }

              \vspace*{2mm}
  \caption{A canonical example of a single address reduction.}
  \label{fig:array_sum_code}
\end{figure}

Transformations as described above have been the main interest of reduction
handling approaches outside the polyhedral world. Associativity and
commutativity properties are used to extract and parallelize the reduction
loop~\cite{Jouvelot1986,Fisher1994} or to parallelize the reduction computation
with regards to an existing surrounding
loop~\cite{Pinter1991,Pottenger1995,Rauchwerger1995,Suganuma1996,Yu2004}.  While
prior work on reductions in the polyhedral
model~\cite{Redon1993,Redon1994,Redon2000,Gupta2002,Zou2012} was focused on
system of affine recurrences (SAREs), we look at the problems a production
compiler has to solve when we allow polyhedral optimizations that exploit the
reduction properties. To this end our work supplements the polyhedral optimizer
\polly~\cite{Grosser12}, part of the \llvm~\cite{Lattner2004} compiler framework
with awareness for the associativity and commutativity of reduction
computations. While we are still in the process of upstreaming, most parts are
already accessible in the public code repository.

The contributions of this paper include:
\begin{itemize}
  \item A powerful algorithm to identify reduction dependences, applicable
        whenever memory or value based dependence information is available.
  \item A sound model to relax memory dependences with regards to reductions and
        its use in reduction-enabled polyhedral scheduling.
  \item A dependence based approach to identify vectorization and
        parallelization opportunities in the presence of reductions.
\end{itemize}

The remainder of this paper is organized as follows:
We give a short introduction into the polyhedral model in
Section~\ref{sec:ThePolyhedralModel}. Thereafter, in
Section~\ref{sec:DetectingReductions}, our reduction detection is
described. Section~\ref{sec:ParallelExecution} discusses the
benefits and drawbacks of different reduction parallelization schemes, including
privatization. Afterwards, we present different approaches to utilize the
reduction properties in a polyhedral optimizer in
Section~\ref{sec:ModelingReductions}. In the end we evaluate our work
(Section~\ref{sec:Evaluation}), compare it to existing approaches
(Section~\ref{sec:RelatedWork}) and conclude with possible extensions in
Section~\ref{sec:ConclusionAndFutureWork}.

 \section{The Polyhedral Model} \label{sec:ThePolyhedralModel}

The main idea behind polyhedral loop nest optimizations is to abstract from
technical details of the target program. Information relevant to the
optimization goal is represented in a very powerful mathematical model and the
actual optimizations are well understood transformations on this
representation. In the context of optimization for data locality or parallelism,
the relevant information is the iteration space of each statement, as well as
the data dependences between individual statement instances.

\begin{figure}[htbp]
  \centering
  \begin{BVerbatim}[commandchars=\\\{\}]
  \PY{k}{for} \PY{p}{(}\PY{n}{i} \PY{o}{=} \PY{l+m+mi}{0}\PY{p}{;} \PY{n}{i} \PY{o}{\PYZlt{}} \PY{n}{NX}\PY{p}{;} \PY{n}{i}\PY{o}{+}\PY{o}{+}\PY{p}{)} \PY{p}{\PYZob{}}
\PY{n+nl}{R:}  \PY{n}{q}\PY{p}{[}\PY{n}{i}\PY{p}{]} \PY{o}{=} \PY{l+m+mi}{0}\PY{p}{;}
    \PY{k}{for} \PY{p}{(}\PY{n}{j} \PY{o}{=} \PY{l+m+mi}{0}\PY{p}{;} \PY{n}{j} \PY{o}{\PYZlt{}} \PY{n}{NY}\PY{p}{;} \PY{n}{j}\PY{o}{+}\PY{o}{+}\PY{p}{)} \PY{p}{\PYZob{}}
\PY{n+nl}{S:}    \PY{n}{q}\PY{p}{[}\PY{n}{i}\PY{p}{]} \PY{o}{=} \PY{n}{q}\PY{p}{[}\PY{n}{i}\PY{p}{]} \PY{o}{+} \PY{n}{A}\PY{p}{[}\PY{n}{i}\PY{p}{]}\PY{p}{[}\PY{n}{j}\PY{p}{]} \PY{o}{*} \PY{n}{p}\PY{p}{[}\PY{n}{j}\PY{p}{]}\PY{p}{;}
\PY{n+nl}{T:}    \PY{n}{s}\PY{p}{[}\PY{n}{j}\PY{p}{]} \PY{o}{=} \PY{n}{s}\PY{p}{[}\PY{n}{j}\PY{p}{]} \PY{o}{+} \PY{n}{r}\PY{p}{[}\PY{n}{i}\PY{p}{]} \PY{o}{*} \PY{n}{A}\PY{p}{[}\PY{n}{i}\PY{p}{]}\PY{p}{[}\PY{n}{j}\PY{p}{]}\PY{p}{;}
    \PY{p}{\PYZcb{}}
  \PY{p}{\PYZcb{}}

\end{BVerbatim}
   \caption{BiCG Sub Kernel of BiCGStab Linear Solver.}
  \label{fig:bicg_code}
\end{figure}
\vspace*{-2mm}

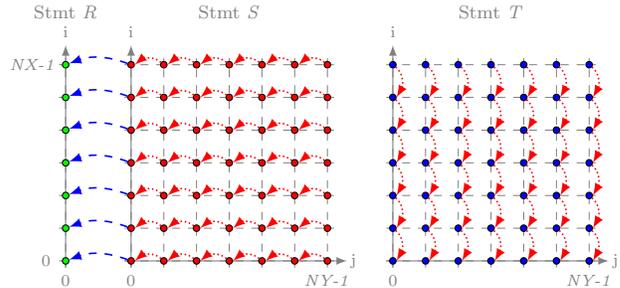
\begin{figure}[htbp]
  \def \scalingfactor {0.87}
\centering

\begin{tikzpicture}[
  scale=\scalingfactor,
  every node/.style={scale=\scalingfactor},
  label/.style={font=\scriptsize,color=gray},
  stmt/.style={font=\small,color=gray},
  axis/.style={thin,gray,-latex},
  grid/.style={style=help lines,dashed},
  inst/.style={draw,circle,inner sep=1pt,fill},
  r/.style={fill=green},
  s/.style={fill=red},
  t/.style={fill=blue},
  dep/.style={semithick,dashed,blue,->},
  lcdep/.style={semithick,densely dotted,red,->},
    >={latex},
  ]
  \draw [axis] (0,-.1) -- (0,3.3);
  \draw[grid] (-.1, -.1) grid[step=.5cm] (.1,3.1);

  \foreach \i in {0,1,...,6} {
    \node[inst,r] (r\i) at (0,\i*0.5) {};
  }

  \node[label] at (0,3.5) {i};
  \node[label] at (-.3,0) {0};
  \node[label] at (0,-.3) {0};
  \node[label] at (-.5,3) {\emph{NX-1}};

  \node[stmt] at (0, 3.8) {Stmt \emph{R}};

  \pgftransformxshift{1cm}

  \draw[axis] (-.1,0) -- (3.3,0);
  \draw[axis] (0,-.1) -- (0,3.3);
  \draw[grid] (-.1, -.1) grid[step=.5cm] (3.1,3.1);

  \foreach \i in {0,1,...,6} {
    \node[inst,s] (s\i0) at (0,\i*0.5) {};
    \draw [dep] (s\i0) edge[out=160,in=20] (r\i);
    \foreach \j in {1,2,...,6} {
      \pgfmathsetmacro\lastj{\j - 1}
      \node[inst,s] (s\i\j) at (\j*0.5,\i*0.5) {};
      \draw [lcdep] (s\i\j) edge[out=140,in=40] (s\i\lastj);
    }
  }

  \node[label]
  at (0,3.5) {i};
  \node[label] at (3.4,0) {j};
  \node[label] at (0,-.3) {0};
  \node[label] at (3,-.3) {\emph{NY-1}};

  \node[stmt] at (1.5, 3.8) {Stmt \emph{S}};

  \pgftransformxshift{4cm}

  \draw [axis] (-.1,0) -- (3.3,0);
  \draw [axis] (0,-.1) -- (0,3.3);
  \draw[grid] (-.1, -.1) grid[step=.5cm] (3.1,3.1);

  \foreach \j in {0,1,...,6} {
    \node[inst,t] (t\j0) at (\j*0.5, 0) {};
    \foreach \i in {1,2,...,6} {
      \pgfmathsetmacro\lasti{\i - 1}
      \node[inst,t] (t\j\i) at (\j*0.5,\i*0.5) {};
      \draw [lcdep] (t\j\i) edge[out=-40,in=60] (t\j\lasti);
    }
  }

  \node[label] at (0,3.5) {i};
  \node[label] at (3.4,0) {j};
  \node[label] at (0,-.3) {0};
  \node[label] at (3,-.3) {\emph{NY-1}};

  \node[stmt] at (1.5, 3.8) {Stmt \emph{T}};
\end{tikzpicture}
   \caption{Polyhedral representation of statements \emph{R}, \emph{S} and
  \emph{T} of the BiCG Sub Kernel of Figure~\ref{fig:bicg_code}.}
                        \label{fig:bicg_polyhedral}
\end{figure}

Figure~\ref{fig:bicg_code} shows an example program containing three statements
\emph{R}, \emph{S} and \emph{T} in a loop nest of depth two.
Figure~\ref{fig:bicg_polyhedral} shows the polyhedral representation of the
individual iteration spaces for all statements, as well as value-based data
dependences
between individual instances thereof. \emph{R} has a one-dimensional iteration
space, as it is nested in the \emph{i}-loop only.  Statements \emph{S} and
\emph{T} have a two-dimensional iteration space as they are nested in both the
\emph{i}-loop as well as in the \emph{j}-loop.  The axes in the Figure
correspond to the respective loops.  Single instances of each statement are
depicted as dots in the graph. Dependences between individual statement
instances are depicted as arrows: dashed ones for regular data dependences and
dotted ones for loop carried data dependences.

\vspace*{-2mm}

In the polyhedral model the iteration space of a statement \emph{Q} is
represented as a multidimensional \(\mathbb{Z}\)-polytope \(\mathcal{I}_Q\),
defined by affine constraints on the iteration variables of loops surrounding
the statement, as well as global parameters. The latter are basically loop
invariant expressions like for example the upper bounds \texttt{NX} and
\texttt{NY} of the loops in Figure~\ref{fig:bicg_code}. As a consequence, the
polyhedral model is only applicable to well structured program parts with affine
loop bounds and memory access functions, so called \textit{Static Control Parts}
(\pscops)~\cite{Grosser12}. While, there are different over-approximations to
increase the applicability (e.g., by Benabderrahmane~\cite{Benabderrahmane2010})
we will assume that all restrictions of \pscops are fulfilled.

The dependences between two statements \emph{Q} and \emph{T} are also
represented as a multidimensional \(\mathbb{Z}\)-polytope known as the
dependence polytope \(\mathcal{D}_{<Q,T>}\). It contains a point \(<i_Q, i_T>\)
for every pair of instances \(<i_Q>\in\mathcal{I}_Q\) and
\(<i_T>\in\mathcal{I}_T\) for which the latter depends on the former.
To ease reading we will however omit the index of the dependence polytopes and
only argue about the set of all dependences \DepSet, defined as:
\[ \DepSetMath := \{<i_Q, i_T>\ |\ \forall Q,T \in \pscop: <i_Q, i_T> \in \DepSetMath_{<Q,T>}\}\]
Later we will also distinguish all \textit{Write-After-Write} (\textit{WAW} or
output dependence) dependences of \DepSet by writing
$\DepSetMath_{\mathit{WAW}}$.

A loop transformation in the polyhedral model is represented as an affine
function \(\theta_Q\) for each statement \emph{Q}. It is often called scheduling
or scattering function. This function translates a point in the original
iteration space \(\mathcal{I}_Q\) of statement \emph{Q} into a new, transformed
target space. One important legality criterion for such a transformation is that
data dependences need to be respected: The execution of every instance of a
source statement \emph{Q} of a dependence has to precede the execution of the
corresponding target statement \emph{T} in the transformed space. Formulated
differently: the target iteration vector of the value producing instance of
\emph{Q} has to be lexicographically smaller\footnote{To compare two vectors of
  different dimensionality, we simply fill up the shorter vector with zeros in
  the end to match the dimensionality of the larger one.} than the target
  iteration vector of the consuming instance of \emph{T}:
\begin{align}
  <i_Q, i_T> \in \DepSetMath \Rightarrow \theta_Q(i_Q) \ll \theta_T(i_T)
  \label{eq:CausalityCondition}
\end{align}
Multiple statements, or multiple instances of the same statement, that are
mapped to the same point in the target space, can be executed in parallel.
However, implementations of polyhedral
schedulers~\cite{Bondhugula2008,Verdoolaege2010} usually generate
scheduling functions with full rank, thus \(\mathit{rank}(\mathit{dom}(\theta)) =
\mathit{rank}(\mathit{img}(\theta))\). The parallelism is therefore not explicit in the scheduling
function but is exposed later when the polyhedral representation is converted to
target code.

There are two things that make the described model particularly interesting for
loop transformation: First, unlike classical optimizers, a polyhedral optimizer
does not only consider individual statements, but instead individual dynamic
instances of each statement.  This granularity leads to a far higher
expressiveness. Second, the combination of multiple classical loop
transformations, like for instance loop skewing, reversal, fusion, even tiling,
typically used as atoms in a sequence of transformations, can be performed in
one step by the scattering function. There is no need to come up with and
evaluate different, possibly equivalent or even illegal combinations of
transformations.  Instead, linear optimization is used to optimize the
scattering function for every individual statement with respect to an
optimization goal.

\begin{figure*}[htbp]
  \centering
  \begin{tabular}{lccl}
    \tf({\tt l = load A[f($i_S, p$)]})  &$:=$\hspace*{-0mm}& $\lambda l:$& if $(l\ \neq\ \texttt{l})$ then $\bot$ else\vspace*{0.15cm}\\
                                        &    \hspace*{-0mm}&             & if $(\mathit{hasOutsideUses}(l))$ then $\top$ else $\uparrow$ \vspace*{0.15cm}\\
    \tf({\tt x = y $\odot$ z}) &$:=$\hspace*{-0mm}& $\lambda l:$& if $(\{\,(\tf(y))(l),\: (\tf(z))(l)\,\} = \{\,\bot, \bot\,\})$  then $\bot$ else\vspace*{0.15cm}\\
                               &   \hspace*{-0mm}&             & if $\neg(\mathit{isCommutative}(\odot) \land \mathit{isAssociative}(\odot)) $\hspace*{0.8mm} then $\top$ else \vspace*{0.15cm}\\
                               &   \hspace*{-0mm}&             & if $(\mathit{hasOutsideUses}(x))$\hspace*{0.8mm} then $\top$ else\vspace*{0.15cm}\\
                               &   \hspace*{-0mm}&             & if $(\{\,(\tf(y))(l),\: (\tf(z))(l)\,\} = \{\,\uparrow, \bot\,\})$\hspace*{0.8mm} then $\odot$ else\vspace*{0.15cm}\\
                               &   \hspace*{-0mm}&             & if $(\{\,(\tf(y))(l),\: (\tf(z))(l)\,\} = \{\,\odot, \bot\,\})$  then $\odot$ else $\top$ \vspace*{0.15cm}\\
    \tf({\tt store x A[f($i_S, p$)]})     &$:=$\hspace*{-0mm}& $\lambda l:$& if $\texttt{x}\ \notin\ $\Insts then $\bot$ else \vspace*{0.15mm}\\
                                          &    \hspace*{-0mm}&             & if $(\mathit{ran}(l)\ \cap\ ran(\texttt{A[f($i_S, p$)]}) = \emptyset)$ then $\top$ else \vspace*{0.15mm}\\
                                &    \hspace*{-0mm}&             & if $(\exists l': l \neq l' \land ran(l') \cap ran(l) \cap ran(\texttt{A[f($i_S, p$)]}) \neq \emptyset)$ then $\top$ else (\tf(\texttt{x}))($l$)\\
                                & & &\\
  \end{tabular}
  \caption{Detection function for reduction-like computations: \tf: \Insts \(\to (\latticeL \to
  \latticeO)\).}
  \label{fig:transformer_function}
\end{figure*}

\section{Detecting Reductions}\label{sec:DetectingReductions}Pattern based approaches on source statements are limited to
find general reduction idioms~\cite{Fisher1994,Rauchwerger1995,Suganuma1996}.
The two main restrictions are the amount of patterns in the compiler's reduction
pattern database and the sensitivity to the input code quality or preprocessing
steps. To become as independent as possible of source code quality and
canonicalization passes we replace the pattern recognition by a simple,
data flow like analysis. This analysis will identify \textit{reduction-like
computations} within each polyhedral statement. Such a computation is a
potential candidate for a reduction, thus it might be allowed to perform the
computation in any order or even in parallel. Afterwards, we utilize
the polyhedral dependence analysis~\cite{Feautrier91} to precisely identify all
reduction dependences~\cite{Pugh1994} in a \pscop, hence to identify the actual reduction
computations from the set of possible candidate (reduction-like) computations.
\begin{figure}[htbp]
  \centering
  \tt l = load A[f($i_S, p$)] \hspace*{4mm} \tt store x A[f($i_S, p$)] \\
  \tt x = y $\odot$ z \\
    \vspace*{2mm}
  \caption{SSA-based language subset.}
  \label{fig:language}
\end{figure}

The following discussion is restricted to the SSA-based
language subset (\Insts) depicted in Figure~\ref{fig:language}. Our
implementation however handles all LLVM-IR~\cite{Lattner2004} instructions.

The binary operation is parametrized with $\odot$ and can be instantiated with
any arithmetic, bit-wise or logic binary operator. To distinguish associative
and commutative binary operators we use $\oplus$ instead.  The \loadI
instruction is applied to a memory location. It evaluates to the current value
{\tt x} stored in the corresponding memory location.  The {\tt store}
instruction takes a value {\tt x} and writes it to the given memory location. In
both cases the memory location is described as {\tt A[f($i_S, p$)]}, where {\tt
A} is a constant array base pointer and {\tt f($i_S, p$)} is an affine function
with regards to outer loop indices of the statement \(S\) ($i_S$) and parameters
($p$) of the \pscop. The range of a memory instruction is defined as the range
of its affine access function:
\begin{align*}
  \mathit{ran}(\text{\tt store x A[f($i_S, p$)]}) &:=\ \mathit{ran}(\text{\tt A[f($i_S, p$)]})
  \\ \mathit{ran}(\text{\tt load A[f($i_S, p$)]}) &:=\ \mathit{ran}(\text{\tt A[f($i_S, p$)]})
  \\ \mathit{ran}(\text{\tt A[f($i_S, p$)]})       &:=\ \text{\tt A} +
  \mathit{ran}(\text{\tt f($i_S, p$)})
\end{align*}

Note the absence of any kind of control flow producing or dependent
instructions ($\phi$~instructions or branches). This is a side effect of the
limited scope of the reduction detection analysis. It is applied only to
polyhedral statements, in our setting basic blocks with exactly one \storeI
instruction. Furthermore, we assume all loop carried values are communicated
in memory.
This setup is equivalent to C source code statements
without non-memory side effects.

\subsection{Reduction-like Computations}\label{sec:ReductionLikeComputations}Reduction-like computations are a generalization of the reduction definition
used e.g., by Jouvelot~\cite{Jouvelot1989} or
Rauchwerger~\cite{Rauchwerger1995}. Their main characteristic is an associative
and commutative computation which reduces a set of input values into reduction
locations. Furthermore, the input values, the control flow and any value that
might escape into a non-reduction location needs to be independent of the
intermediate results of the reduction-like computation. The difference between
reduction-like computations and reductions known in the literature is the
restriction on other appearances of the reduction location in the loop nest.
We do not restrict syntactic appearances of the reduction location base
pointer as e.g., Rauchwerger~\cite{Rauchwerger1995} does, but only accesses to
the actual reduction location in the same statement. This means a
reduction-like computation on \texttt{A[i\%2]} is not invalidated by any
    occurrence of \texttt{A[i\%2 + 1]} in the same statement or any occurrence
        of \texttt{A} in another statement.

It is crucial to stress that we define reduction-like computations for a
single polyhedral statement containing only a single \storeI. Thus intermediate
results of a reduction-like computation can only escape if they are used in a
different statement or outside the \pscop. As we focus on memory
reductions in a single statement we will assume such outside uses invalidate a
candidate computation from being reduction-like. To this end we define the
function:
\vspace*{-1mm}\[\mathit{hasOutsideUses}: \text{\Insts} \to bool
\vspace*{-1mm}\]

that returns true if an instruction is used outside its statement. In
Section~\ref{sec:GeneralPolyhedralStatements} we explain how the situation
changes if multiple statements are combined into \textit{compound statements} in
order to save compile time.

Reconsider the array sum example in Figure~\ref{fig:array_sum_code}. The
reduction location is the variable {\tt sum}, a scalar variable or zero
dimensional array. However, we do not limit reduction-like computations to zero
dimensional reduction locations, instead we allow multidimensional reduction
locations, also called \textit{histogram reductions}~\cite{Pottenger1995}, as
well. The second example, Figure~\ref{fig:bicg_code}, shows two such
multidimensional reductions. The reduction locations are {\tt q[i]} and {\tt
s[j]}. The first is variant in the outer loop, the second in the inner loop.

To detect reduction-like computations we apply the detection function \tf, shown
in Figure~\ref{fig:transformer_function}, to the  \storeI in the polyhedral
statement \(S\). The idea is to track the flow of
loaded values through computation up to the \storeI. To this end, \tf({\tt I})
for any instruction {\tt I} will assign each \loadI a symbol that describes how
the value loaded by \loadI used up to and by {\tt I}. We will use \latticeO\ to
refer to the set of all four symbols. It includes the $\bot$
indicating that the \loadI was not used by the instruction, the $\uparrow$ to
express that it was only loaded but not yet used in any computation, the $\top$
stating that the loaded value may have been used in a non-associative or
non-commutative computation. Additionally, the $\oplus$ is used when the loaded
value was exactly one input of a chain of $\oplus$ operations. Note that only a
\loadIl that flows with $\oplus$ into the \storeI is a valid candidate for a
reduction-like computation and only if the load and the \storeI access
(partially) the same memory.  Furthermore, we forbid all other \loadI
instructions in the statement to access the same memory as both {\tt l} and the
\storeI as that would again make the computation potentially non-associative and
non-commutative.

If a valid \loadIl was found, it is the unique \loadI instruction inside the
statement \(S\) that accesses (partially) the same memory as the \storeIs and
(\tf({\tt s}))({\tt l}) is an associative and commutative operation $\oplus$.
We will refer to the quadruple (\(S\), {\tt l},
$\oplus$, {\tt s}) as the reduction-like computation \redLike\ of \(S\) and
denote the set of all reduction-like computations in a \pscop as \redLikes.

It is worth noting that we explicitly allow the access functions of the \loadI
and the \storeI to be different as for example shown in
Figure~\ref{fig:partial_reduction_code}. In such cases a reduction can manifest
only for certain parameter valuations or, as shown, for certain valuations of
outer loop indices. Additionally, we could easily extend the definition to
allow non-affine but Presburger accesses
or even over-approximated non-affine accesses
if they are pure. It is also worth to note that our definition does not restrict the
shape of the induced reduction dependences.

\begin{figure}[htbp]
  \centering
    \hspace*{1cm}
  \begin{BVerbatim}[commandchars=\\\{\}]
\PY{k}{for} \PY{p}{(}\PY{n}{i} \PY{o}{=} \PY{l+m+mi}{0}\PY{p}{;} \PY{n}{i} \PY{o}{\PYZlt{}} \PY{n}{N}\PY{p}{;} \PY{n}{i}\PY{o}{+}\PY{o}{+}\PY{p}{)}
  \PY{k}{for} \PY{p}{(}\PY{n}{j} \PY{o}{=} \PY{l+m+mi}{0}\PY{p}{;} \PY{n}{j} \PY{o}{\PYZlt{}} \PY{n}{M}\PY{p}{;} \PY{n}{j}\PY{o}{+}\PY{o}{+}\PY{p}{)}
    \PY{n}{A}\PY{p}{[}\PY{n}{j}\PY{p}{]} \PY{o}{=} \PY{n}{A}\PY{p}{[}\PY{n}{j}\PY{o}{\PYZhy{}}\PY{n}{i}\PY{p}{]} \PY{o}{+} \PY{n}{Mat}\PY{p}{[}\PY{n}{i}\PY{p}{]}\PY{p}{[}\PY{n}{j}\PY{p}{]}\PY{p}{;}

\end{BVerbatim}
   \hspace*{1cm}
  \caption{Conditional reduction with different access functions.}
  \label{fig:partial_reduction_code}

              \end{figure}

\subsection{Reduction Dependences}
\label{sec:ReductionDependences}
While the data flow analysis performed on all polyhedral statements
only marks reduction-like computations, we are actually interested in
{\it reduction dependences}~\cite{Pugh1994}. These loop carried self dependences
start and end in two instances of the same reduction-like computation and they
inherit some properties of this computation. Similar to the reduction-like
computation, reduction dependences can be considered to be ``associative'' and
``commutative''. The latter allows a schedule to reorder the iterations
participating in the reduction-like computation while it can still be considered
valid, however all non-reduction dependences still need to be fulfilled.

We split the set of all dependences \DepSet into the set of reduction
dependences
\DepSetRed $\subseteq$ \DepSet and the set of non-reduction dependences
\DepSetNonRed $:=$ \DepSet $\setminus$ \DepSetRed. Now we can express the
commutativity of a reduction dependence by extending the causality condition
given in Constraint~\ref{eq:CausalityCondition} as follows:
\begin{align}
  <i_Q, i_T> \in \DepSetNonRedMath\ \Longrightarrow\ \theta_Q(i_Q) \ll
  \theta_T(i_T)
  \label{eq:CausalityConditionNonReduction} \\
  <i_Q, i_T> \in \DepSetRedMath\ \Longrightarrow\ \theta_Q(i_Q) \neq
  \theta_T(i_T)
  \label{eq:CausalityConditionReduction}
\end{align}

Constraint~\ref{eq:CausalityConditionNonReduction} is the same as the
original causality condition (Constraint~\ref{eq:CausalityCondition}),
except that we restrict the domain to non-reduction dependences \DepSetNonRed. For
the remaining reduction dependences
\DepSetRed, Constraint~\ref{eq:CausalityConditionReduction} states that the
schedule $\theta$ can reorder two iterations freely, as long as they are not
mapped to the same time stamp. However, relaxing the causality
condition for reduction dependences is only valid if \DepSet contains all
transitive reduction dependences. This is for example the case if \DepSet is
computed by a memory-based dependence analysis. In case only value-based
dependence analysis~\cite{Feautrier91} was performed it is also sufficient to
provide the missing transitive reduction dependences e.g., by recomputing them
using a memory-based dependence analysis.

Reconsider the BiCG kernel (Figure~\ref{fig:bicg_code}) and its non transitive
(value-based) set of dependences \DepSet shown in
Figure~\ref{fig:bicg_polyhedral}. If we remove all reduction dependences
\DepSetRed from \DepSet, the only constraints left involve statement \(R\) and
the iterations of statements \(S\) with \(j = 0\). Consequently, there is no
reason not to schedule the other instances of statement \(S\) before statement
\(R\).

To address the issue of only value-based dependences without recomputing
memory-based ones we use the transitive closure \(\DepSetRedMath^{+}{_S}\) of
the reduction dependences for a statement \(S\)
(Equation~\ref{eq:reduction_dependences_transitive}). As the transitive closure
of a Presburger relation is not always a Presburger relation we might have to
use an over-approximation to remain sound, however Pugh and
Wonnacot~\cite{Pugh1991} describe how the transitive closure can also be
computed precisely for exact direction/distance vectors. They also argue in
later work~\cite{Pugh1994} that the transitive closure of value-based reduction
dependences of real programs can be computed in an easy and fast way.

If we now interpret \(\DepSetRedMath^{+}{_S}\) as a relation that maps instances
of a reduction statement \(S\) to all instances of \(S\) transitively dependent,
we can define \textit{privatization dependences} \DepSetPriv
(Equation~\ref{eq:privatization_dependences}).  In simple terms, \DepSetPriv
will ensure that no non-reduction statement accessing the reduction location can
be scheduled in-between the reduction statement instances by extending the
dependences ending or starting from a reduction access to all reduction access
instances. This also means that in case no memory locations are reused e.g.,
after renaming and array expansion~\cite{Feautrier88} was applied, the set of
privatization dependences will be empty.
\begin{align}
  \DepSetRedMath^{+}{_S} &:=\ (\DepSetRedMath\ \cap <\mathcal{I}_S, \mathcal{I}_S>)^+
  \label{eq:reduction_dependences_transitive} \\
  \DepSetPrivMath &:=\ \{<i_T, \DepSetRedMath^{+}{_S}(i_{S})>\ |\ <i_T, i_S> \in \DepSetMath_{<T, S>} \}\notag\\
                  &\hspace*{1.8mm} \cup\ \{<\DepSetRedMath^{+}{_S}(i_{S}), i_T>\ |\ <i_S, i_T> \in \DepSetMath_{<S, T>} \} \label{eq:privatization_dependences}
\end{align}

Privatization dependences overestimate the dependences that manual privatization
of the reduction locations would cause. They are used to create alternative
causality constraint for the reduction statements that enforce the initial order
between the reduction-like computation and any other statement accessing the
reduction locations. To make use of them we replace
Constraint~\ref{eq:CausalityConditionNonReduction} by
Constraint~\ref{eq:CausalityConditionNonReductionPriv}.
\begin{align}
  <i_Q, i_T> \in (\DepSetNonRedMath \cup \DepSetPrivMath)\ \Longrightarrow\ \theta_Q(i_Q) \ll \theta_T(i_T)
  \label{eq:CausalityConditionNonReductionPriv}
\end{align}

If we now utilize the associativity of the reduction dependences we can compute
intermediate results in any order before we combine them to the final result.
As a consequence we can allow parallelization of the reduction-like computation,
thus omit Constraint~\ref{eq:CausalityConditionReduction}; {\it thereby
  eliminating the reduction dependences \DepSetRed from the causality condition
of a schedule completely}.  However, parallel execution of iterations connected
by reduction dependences requires special ``treatment'' of the accesses during
the code generation as described in Section~\ref{sec:ParallelExecution}.

The restriction on polyhedral statements, especially that it contains at most
one \storeI instruction, eases the identification of reduction dependences;
they are equal to all loop carried \textit{Write-After-Write} self dependences
over a statement with a reduction-like computation\footnote{In this restricted
environment we could also use the \textit{Read-After-Write} (RAW) dependences
instead of the WAW ones.}. Thus, \DepSetRed can be expressed as stated in
Equation~\ref{eq:reduction_dependences}.
\begin{align}
\text{\DepSetRed} := \text{\DepSet}_{\textit{WAW}} \cap \{\ \mathcal{I}_S \times \mathcal{I}_S \ |\ (S, {\tt l},
  \oplus, {\tt s}) \in \redLikes \}
  \label{eq:reduction_dependences}
\end{align}

\subsection{General Polyhedral Statements}
\label{sec:GeneralPolyhedralStatements} Practical polyhedral optimizer operate
on different granularities of polyhedral statements; a crucial factor for both
compile time and quality of the optimization.  While
\textit{Clan}~\footnote{http://icps.u-strasbg.fr/\~{}bastoul/development/clan/}
operates on C statements, \polly is based on basic blocks in the SSA-based
intermediate language of LLVM.  The former eases not only reduction handling but
also offers more scheduling freedom. However, the latter can accumulate the
effects of multiple C statements in one basic block, thus it can perform better
with regards to compile time.  Finding a good granularity for a given program,
e.g., when and where to split a LLVM basic block in the Polly setting, is a
research topic on its own but we do not want to limit our approach to one fixed
granularity. Therefore, we will now assume a polyhedral statement can contain
multiple \storeI instructions, thus we allow arbitrary basic blocks.

As a first consequence we have to check that intermediate values of a
reduction-like computation do not escape into non-reduction memory locations.
This happens if and only if intermediate values---and therefore the reduction
\loadI---flow into multiple \storeI instructions of the statement \(S\).
Additionally, other \storeI instructions are not allowed to override
intermediate values of the reduction computation. Thus, \((S,\text{\tt
l},\oplus,\text{\tt s})\) can only be a reduction-like computations, if for all
other \storeI instructions \texttt{s'} in \(S\): \[(t(\text{\tt s'}))(\text{\tt
  l}) = \bot\ \land\ range(\texttt{s'})
\cap range(\texttt{s}) \cap range(\texttt{l}) = \emptyset\]

Furthermore, we cannot assume that all loop carried \textit{WAW} self
dependences of a statement containing a reduction-like computation are reduction
dependences: other read and write accesses contained in the statement could
cause the same kind of dependences. However, we are particularly interested in
dependences caused by the \loadI and \storeI instruction of a reduction-like
computation \redLike. To track these accesses separately we can pretend they are
contained in their own statement \(S_{\redLike}\) that is executed at the same
time as \(S\) (in the original iteration space). This is only sound as long as
no other instruction in \(S\) accesses (partially) the same memory as the \loadI
or the \storeI, but this was already a restriction on reduction-like
computations. The definition of reduction dependences
(Equation~\ref{eq:reduction_dependences}) is finally changed to: \begin{align}
  \text{\DepSetRed} := \text{\DepSet}_{\mathit{WAW}} \cap \{\ \mathcal{I}_{S_{\redLike}} \times
\mathcal{I}_{S_{\redLike}} \ |\ \redLike \in \redLikes \}
\label{eq:reduction_dependences_general} \end{align}

It is important to note the increased complexity of the dependence detection
problem when we model reduction accesses separately. However, our experiments in
Section~\ref{sec:Evaluation} show that the effect is (in most cases)
negligible. Furthermore, we want to stress that this kind of separation is not
equivalent to separating the reduction access at the statement level as we do
not allow separate scheduling functions for \(S\) and \(S_{\redLike}\). Similar
to a fine-grained granularity at the statement level, separation might be
desirable in some cases, however it suffers from the same drawbacks.
 \section{Parallel Execution}
\label{sec:ParallelExecution}
When executing accesses to a reduction location \|x|, \|p| times in
parallel, it needs to be made sure that the read-modify-write cycle on \|x|
happens atomically. While doing exactly that --- performing atomic
read-modify-write operations --- might be a viable solution in some
contexts~\cite{Yu2004}, it is generally too expensive. The
overhead of an atomic operation easily outweighs the actual work for smaller
tasks~\cite{Pottenger1995}. Additionally, the benefit of vectorization is lost
for the reduction, as atomic operations have to scalarize the computation again.
We will therefore focus our discussion and the evaluation on privatization as it
is generally well-suited for the task at hand~\cite{Pottenger1995}.

\begin{figure}
  \centering
        \centering
    \begin{BVerbatim}[commandchars=\\\{\}]
\PY{c+c1}{// (A) init}
\PY{k}{for} \PY{p}{(}\PY{n}{i} \PY{o}{=} \PY{l+m+mi}{0}\PY{p}{;} \PY{n}{i} \PY{o}{\PYZlt{}} \PY{n}{NX}\PY{p}{;} \PY{n}{i}\PY{o}{+}\PY{o}{+}\PY{p}{)}
  \PY{c+c1}{// (B) init}
  \PY{k}{for} \PY{p}{(}\PY{n}{j} \PY{o}{=} \PY{l+m+mi}{0}\PY{p}{;} \PY{n}{j} \PY{o}{\PYZlt{}} \PY{n}{NY}\PY{p}{;} \PY{n}{j}\PY{o}{+}\PY{o}{+}\PY{p}{)}
    \PY{c+c1}{// (C) init}
    \PY{k}{for} \PY{p}{(}\PY{n}{k} \PY{o}{=} \PY{l+m+mi}{0}\PY{p}{;} \PY{n}{k} \PY{o}{\PYZlt{}} \PY{n}{NZ}\PY{p}{;} \PY{n}{k}\PY{o}{+}\PY{o}{+}\PY{p}{)}
      \PY{n}{P}\PY{p}{[}\PY{n}{j}\PY{p}{]} \PY{o}{+}\PY{o}{=} \PY{n}{Q}\PY{p}{[}\PY{n}{i}\PY{p}{]}\PY{p}{[}\PY{n}{j}\PY{p}{]} \PY{o}{*} \PY{n}{R}\PY{p}{[}\PY{n}{j}\PY{p}{]}\PY{p}{[}\PY{n}{k}\PY{p}{]}\PY{p}{;}
    \PY{c+c1}{// (C) aggregate}
  \PY{c+c1}{// (B) aggregate}
\PY{c+c1}{// (A) aggregate}

\end{BVerbatim}

      \caption{Possible privatization locations (\|A|-\|C|) for the reduction over
  \texttt{P[j]}.
}
  \label{fig:PrivPlacement}
\end{figure}

\subsection{Privatizing Reductions}
\label{sec:Privatization}Privatization means that every parallel context \(\mathid{c}_i\),
which might be a thread or just a vector lane, depending on the
kind of parallelization, gets its own private location \(x_i\) for \|x|. In
front of the parallelized loop carrying a reduction dependence \|p|, private
locations \(x_1, \cdots, x_p\) of \|x| are allocated and initialized with the
identity element of the corresponding reduction operation \(\oplus\). Every parallel context \(\mathid{c}_i\) now
non-atomically, and thus cheaply, modifies its very own location \(x_i\). After the
loop, but before the first use of the \(x\), accumulation code needs to join all
locations into \|x| again, thus: \(x := x \oplus x_1 \oplus \cdots \oplus x_p\).

Such a privatization transformation is legal due to the properties of a
reduction operation.  Every possible user of \|x| sees the same result after the
final accumulation has been performed as it would have seen before the
transformation.  Nevertheless we gained parallelism which cannot be exploited
without the reduction properties. It might seem, that the final accumulation of
the locations needs to be performed sequentially, but note that the number of
locations does not necessarily grow with the problem size but instead only with the
maximal number of parallel contexts. Furthermore, accumulation can be done in
logarithmic time by parallelizing the accumulation
correspondingly~\cite{Fisher1994}.

One positive aspect of using privatization to fix a broken reduction dependence
is that it is particularly well-suited for both ways of parallelization
usually performed in the polyhedral context: thread parallelism and
vectorization. For thread parallelism real private locations of the reduction
address are allocated; in case of vectorization, a vector of suitable width is
used.

As described, privatization creates ``copies'' of the reduction location, one
for each instance possibly executed in parallel. While we can limit the number
of private locations (this corresponds to the maximal number of parallel
contexts), we cannot generally bound the number of reduction locations.
Furthermore, the number of necessary locations, as well as the number of times
initialization and aggregation is needed, varies with the placement of the
privatization code.

Consider the example in Figure~\ref{fig:PrivPlacement}. Different possibilities
exist to exploit reduction parallelism: using placement \|C| for the
privatization, the \|k|-loop could be executed in parallel and only \|p|
private copies of the reduction location are necessary.  There is no benefit
in choosing location \|B| as we then need \(p\times\mathid{NY}\) privatization
locations (we have \|NY| different reduction locations modified by the \|j|-loop
and \|p| parallel contexts), but there is no gain in the amount of parallelism
(the \|j|-loop is already parallel). Finally, choosing location \|A| for
privatization might be worthwhile. We still only need \(p\times\mathid{NY}\)
privatized values, but save aggregation overhead: While for location \|C|, \|p|
values are aggregated \(\mathid{NX}\times\mathid{NY}\) times and for location
\|B|, \(p\times\mathid{NY}\) locations are aggregated \|NX| times, for location
\|A|, \(p\times\mathid{NY}\) locations are aggregated only once. Furthermore, the
\|i|-loop can now be parallelized.

In general, a trade-off has to be made between memory consumption, aggregation
time and exploitable parallelism. Finding a good placement however is difficult
and needs to take the optimization goal, the hardware and the workload size
into account.  Furthermore, depending on the scheduling, the choices for
privatization code placement in the resulting code might be limited, which
suggests that the scheduler should be aware of the implications of a chosen
schedule with respect to the efficiency of necessary privatization.

In Section~\ref{sec:bicg_case_study} we discuss the effect of different
placement choices for the BiCG benchmark shown in Figure~\ref{fig:bicg_code}.

\section{Modeling Reductions}
\label{sec:ModelingReductions}As mentioned earlier, the set \DepSet of all dependences is partitioned into the
set \DepSetRed of reduction induced dependences and \DepSetNonRed of regular
dependences. Reduction dependences inherit properties similar to
associativity and commutativity from the reduction operator $\oplus$: the
corresponding source and target statement instances can be executed in any
order---provided $\oplus$ is a commutative operation---or in parallel---if
$\oplus$ is at least associative. In order to exploit these properties the
polyhedral optimizer needs to be aware of them. To this end we propose different
scheduling and code generation schemes.

\begin{description}
  \item {\it Reduction-Enabled Code Generation}~\\
    is a simple, non-invasive method to realize reductions during the code
    generation, thus without modification of the polyhedral representation of
    the \pscop.
  \item {\it Reduction-Enabled Scheduling}~\\
    exploits the properties of reductions in the polyhedral representation. All
    reduction dependences are basically ignored during scheduling, thereby
    increasing the freedom of the scheduler.
  \item {\it Reduction-Aware Scheduling}~\\
    is the representation of reductions and their realization via privatization
    in the polyhedral optimization. The scheduler decides when and where to make
    use of reduction parallelism. However, non-trivial modifications of the
    polyhedral representation and the current state-of-the-art schedulers are
    necessary.
\end{description}

\subsection{Reduction-Enabled Code Generation} \label{sec:ReductionEnabledCodeGeneration}
The reduction-enabled code generation is a simple, non-invasive approach to
exploit reduction parallelism. The only changes needed to enable this
technique are in the code generation, thus the polyhedral representation is not
modified. So far, dimensions or loops are marked
parallel if they do not carry any dependences. With regards to reduction
dependences we can relax this condition, hence we can mark non-parallel
dimensions or loops as parallel, provided we add privatization code, if they
only carry reduction dependences. To implement this technique we add one
additional check to the code generation that is executed for each non-parallel
loop of the resulting code that we want to parallelize.  It uses only
non-reduction dependences \DepSetNonRed not \DepSet to determine if
the loop exclusively carries reduction dependences. If so, the reduction
locations corresponding to the broken dependences are privatized and the loop is
parallelized.

Due to its simplicity, it is easily integrable into existing optimizers while
the compile time overhead is reasonably low. However, additional heuristics
are needed. First, to decide if reductions should be realized e.g., if
privatization of a whole array is worth the gain in parallelism. And second,
where the privatization statements should be placed (cf.
Section~\ref{sec:Privatization}). Note that usually the code generator has no,
and in fact should not have any, knowledge of the optimization goal of the
scheduler.

Apart from the need for heuristics, reduction-aware code generation also misses
opportunities to realize reductions effectively. This might happen if the scheduler has no
reason to perform an enabling transformation or the applied transformation even
disabled the reduction. Either way, it is hard to predict the outcome
of this approach.

\subsection{Reduction-Enabled Scheduling} \label{sec:ReductionEnabledScheduling}
In contrast to reduction-aware code generation, which is basically
a post-processing step, reduction-enabled scheduling actually influences the
scheduling processes by eliminating reduction dependences beforehand.
Therefore, the scheduler is (1) unaware of the existence of reductions and their
dependences and (2) has more freedom to schedule statements if they
contain reduction instances. While this technique allows to exploit reductions
more aggressively, there are still disadvantages. First of all, this approach
relies on reduction-aware code generation as a back-end, hence it shares the same
problems.  Second, the scheduler's unawareness of reduction dependences prevents
it from associating costs to reduction realization. Thus, privatization is
implicitly assumed to come for free. Consequently, the scheduler does not prefer
existing, reduction-independent parallelism over reduction parallelism and
therefore may require unnecessary privatization code.

For the BiCG example (Figure~\ref{fig:bicg_code}) omitting the reduction
dependences might not result in the desired schedule if we assume we are only
interested in one level of outermost parallelism\footnote{A reasonable
  assumption for desktop computers or moderate servers with a low number of
parallel compute resources.} and furthermore that the statements \(S\) and \(T\)
have been split prior to the scheduling. In this situation we want to
interchange the outer two loops for the \(T\) statement in order to utilize the
inherent parallelism, not the reduction parallelism. However, without the
reduction dependences the scheduler will not perform this transformation. In
order to decrease the severity of this problem, the reduction dependences can
still be used in the proximity constraints of the
scheduler~\cite{Verdoolaege2010}, thus the scheduler will try to minimize the
dependence distance between reduction iterations and implicitly move them to
inner dimensions. This solves the problem for all Polybench benchmarks with
regards to outermost parallelism, however it might negatively affect
vectorization if e.g., the innermost parallel dimension is always vectorized.

\subsection{Reduction-Aware Scheduling} \label{sec:ReductionAwareScheduling}
Reduction-enabled scheduling results in generally good schedules for our
benchmark set, however resource constraints as well as environment effects, both
crucial to the overall performance, are not represented in the typical
objective function used by polyhedral optimizers. In essence we believe, the
scheduler should be aware of reductions and the cost of their privatization, in
terms of memory overhead as well as aggregation costs. This is especially true
if the scheduler is used to decide which dimensions should be executed in
parallel or if there are tight memory bounds (e.g., on mobile devices).

In Section~\ref{sec:bicg_case_study} we show that the execution environment as
well as the values of runtime parameters are crucial factors in the actual
performance of parallelized code, even more when reductions are involved. While
a reduction-aware scheduler could propose different parallelization schemes for
different execution environments or parameter values, there is more work needed
in order to (1) predict the effects of parallelization and privatization on the
actual platform and to (2) express them as affine constraints in the scheduling
objective function.

\newcommand{\AccessCol}{myred}
\newcommand{\AccessDep}{\ref{fig:CompileTimeAccess}\;access-wise\xspace}
\newcommand{\HybridCol}{mygreen}
\newcommand{\HybridDep}{\ref{fig:CompileTimeHybrid}\;hybrid\xspace}
\newcommand{\StatementCol}{myblue}
\newcommand{\StatementDep}{\ref{fig:CompileTimeStmt}\;statement-wise\xspace}

\section{Evaluation} \label{sec:Evaluation}

We implemented Reduction-Enabled Scheduling
(c.f., Section~\ref{sec:ReductionEnabledScheduling}) in the polyhedral optimizer
\polly and evaluated the effects on compile time and run-time on the Polybench
3.2.  We used an \textit{Intel(R) core i7-4800MQ} quad core machine and the
standard input size of the benchmarks.

Our approach is capable of identifying and modeling all reductions as described
in Section~\ref{sec:DetectingReductions}: in total 52 arithmetic reductions in
30 benchmarks~\footnote{This assumes the benchmarks are compiled with
\textit{-ffast-math}, otherwise reductions over floating point computations are
not detected.}.

As described earlier, our detection virtually splits polyhedral statements
to track the effects of the \loadI and \storeI instructions that participate in
reduction-like computations. As this increases the complexity of the performed
dependence analysis we timed this particular part of the compilation for each of
the benchmarks and compared our \HybridDep dependence analysis to a completely \AccessDep analysis and the
default \StatementDep one. We use the term hybrid because reduction
accesses are tracked separately while other accesses are accumulated on
statement level.

\begin{figure*}[htb]
  \centering
      \hspace*{0.5mm}
      \includegraphics[width=\textwidth]{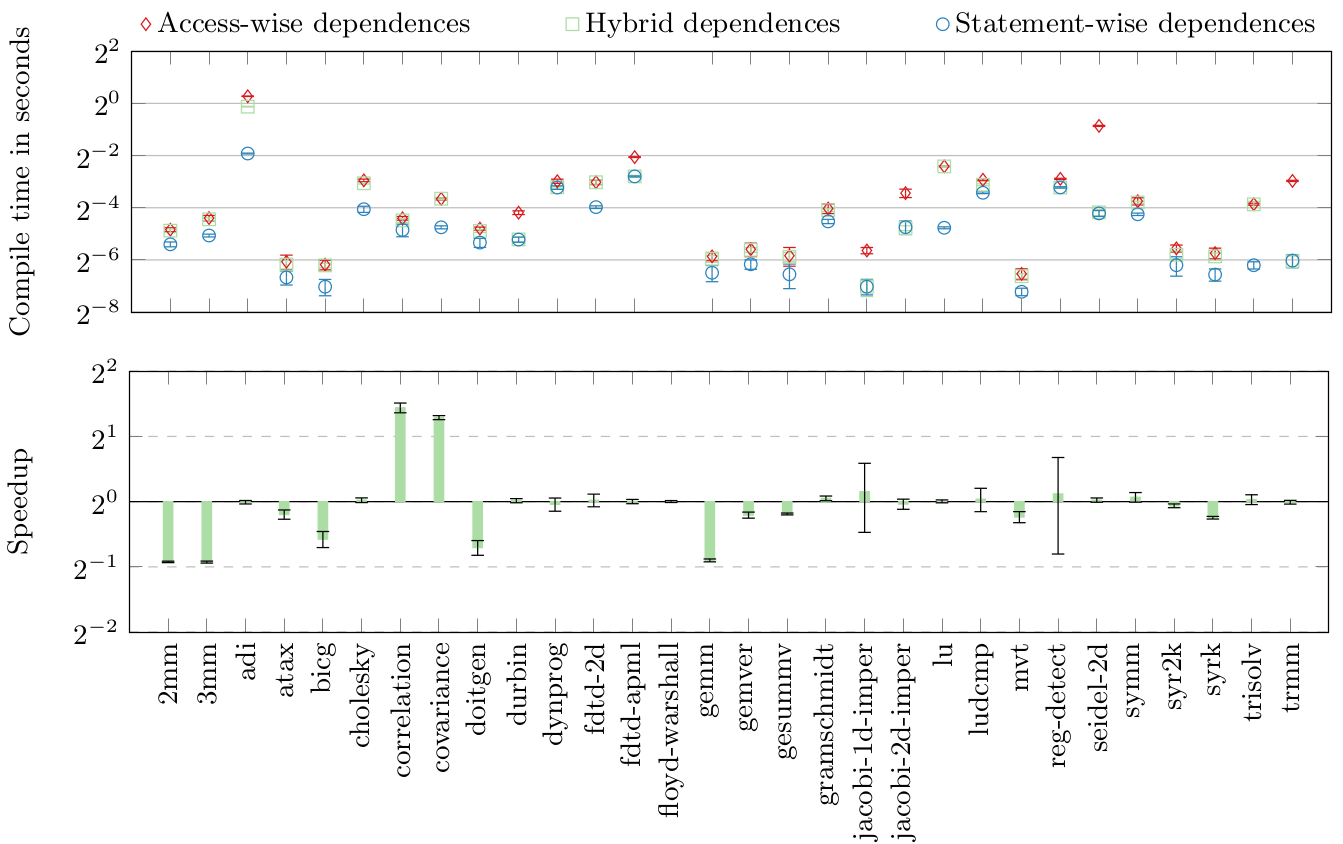}

        \caption{Evaluation results for Polybench 3.2. In the upper part the
      compile time for different grained dependence analyses is shown, in the
      lower part the speedup of Polly with reduction support compared to Polly
      without reduction support.
    }
  \label{fig:Evaluation}
\end{figure*}

As shown in Figure~\ref{fig:Evaluation} (top) our approach takes up to $5\times$
as long (benchmark lu) than the default implementation but in average only 85\%
more.  Access-wise dependence computation however is up to 10$\times$ slower
than the default and takes in average twice as long as our hybrid approach. Note
that both approaches do not only compute the dependences (partially) on the
access level but also the reduction and privatization dependences as
explained in Section~\ref{sec:ReductionDependences}.

Figure~\ref{fig:Evaluation} (bottom) shows the speedup of our approach compared
to the non-reduction \polly. The additional scheduling freedom causes speedups
for the data-mining applications (correlation and covariance) but slowdowns
especially for the matrix multiplication kernels (2mm, 3mm and gemm). This is
due to the way \polly generates vector code.  The deepest dimension of the
new schedule that is parallel (or now reduction parallel) will be strip-mined
and vectorized.  Hence the stride of the contained accesses, crucial to generate
efficient vector code, is not considered.  However, we do not believe this to be
a general shortcoming of our approach as there are existing approaches to tackle
the problem of finding a good vector dimension~\cite{Kong2013} that would
benefit from the additional scheduling freedom as well as the knowledge of
reduction dependences.

\begin{table}
  \centering
  \scriptsize
  \begin{tabular}{l|cc|cc|cc|cc|}
  Parallel    & \multicolumn{2}{c|}{$2^{10}\times2^{10}$} &
  \multicolumn{2}{c|}{$2^{12}\times2^{12}$} &
    \multicolumn{2}{c|}{$2^{14}\times2^{14}$} &
    \multicolumn{2}{c|}{$2^{15}\times2^{15}$}  \\
    \hline
    \textit{Outer} & 0.19&0.55 & 2.31&0.75 & 3.91&0.72 & 2.19&0.96\\
    \textit{Tile}  & 0.03&1.10 & 0.32&1.54 & 0.10&1.60 & 0.16&2.21\\
  \end{tabular}
  \vspace*{2mm}
  \caption{BiCG run-time results. The values are speedups compared to the
    sequential \polly version, first for the 32-core machine, then for the
  4-core machine.
} \label{tbl:BiCG_case_runtime}
\end{table}

\subsection{BiCG Case Study} \label{sec:bicg_case_study}
Polybench is a collection of inherent parallel programs, there is only one---the
BiCG kernel--- that depends on reduction parallelism. To study the effects of
parallelization combined with privatization of multidimensional reductions in the
BiCG kernel we compared two parallel versions to the non-parallel code \polly
would generate without reduction support. The first version ``\textit{Outer}''
has a parallel outermost loop and therefore needs to privatize the whole array
\texttt{s}. The second version ``\textit{Tile}'' parallelizes the second
outermost loop. Due to tiling, only ``tile size'' (here 32) locations of the
\texttt{q} array need to be privatized.  Table~\ref{tbl:BiCG_case_runtime} shows
the speedup compared to the sequential version for both a quad core machine
and a $8\times4$-core server. As the input grows larger the threading overhead
as well as the inter-chip communication on the server will cause the speedup of
\textit{Tile} to stagnate, however on a one chip architecture this version
generally performs best. \textit{Outer} on the other hand will perform well
on the server but not on the 4-core machine.  We therefore believe the
environment is a key factor in the performance of reduction-aware
parallelization and a reduction-aware scheduler is needed to decide under which
run-time conditions privatization becomes beneficial.

\section{Related Work} \label{sec:RelatedWork}

Reduction aware loop parallelization has been a long lasting research topic.
Different approaches to detect reduction, to model them and finally to optimize
them have been proposed. As our work has some intersection with all three parts
we will discuss them in separation.

\subsection{Detection}
Reduction detection started with pattern based approaches on source
statements~\cite{Jouvelot1989,Pinter1991,Redon1993,Pottenger1995,Rauchwerger1995,Redon2000}
and evolved to more elaborate techniques that use symbolic
evaluation~\cite{Fisher1994}, a data dependency graph~\cite{Suganuma1996} or
even a program dependency graph~\cite{Pinter1991} to find candidates for
reduction computations.

For functional programs Xu et al.~\cite{Xu04} use a type system to deduce
parallel loops including pattern based reductions. Their typing rules are
similar to our detection function (Figure~\ref{fig:transformer_function}) we use
to identify reduction-like computations.

Sato and Iwasaki~\cite{Sato2011} describe a pragmatic system to detect and
parallelize reduction and scan operations based on the ideas introduced by
Matsuzaki et al.~\cite{Matsuzaki2006}: the representation of (part of) the loop
as a matrix multiplication with a state vector. They can handle mutually
recursive scan and reduction operations as well as maximum computations
implemented with conditionals, but they are restricted to innermost loops and
scalar accumulation variables. As an extension Zou and Rajopadhye~\cite{Zou2012}
combined the work with the polyhedral model and the recurrence detection
approach of Redon and Feautrier~\cite{Redon1993,Redon2000}. This combination
overcomes many limitations, e.g., multidimensional reductions (and scans) over
arrays are handled. However, the applicability is still restricted to scans and
reductions representable in State Vector Update Form~\cite{Kogge73}.

In our setting we identify actual reductions utilizing the already present
dependence analysis, an approach very similar to the what Suganuma et
al.~\cite{Suganuma1996} proposed to do. However, we only perform the expensive,
access-wise dependence analysis for reduction candidates, and not for all
accesses in the \pscop. Nevertheless, both detections do not need the reductions
to be isolated in a separate loop as assumed by Fisher and
Ghuloum~\cite{Fisher1994} or Pottenger and Eigenmann~\cite{Pottenger1995}.
Furthermore, we allow the induced reduction dependences to be of any form and
carried by any subset of outer loop dimensions. This is similar to the nested
\textit{Recur} operator introduced by Redon and
Feautrier~\cite{Redon1993,Redon2000}. Hence, reductions are not only restricted
to a single loop dimension, as in other
approaches~\cite{Jouvelot1989,Fisher1994,Sato2011}, but can also be
multidimensional as shown in Figure~\ref{fig:bicg_code}.

\subsection{Modeling}
Modeling reductions was commonly done implicitly, e.g., by ignoring the
reduction dependences during a post parallelization
step~\cite{Jouvelot1989,Pinter1991,Redon1993,Pottenger1995,Rauchwerger1995,Xu04,Venkat2014}.
This is comparable to the reduction-enabled code generation described in
Section~\ref{sec:ReductionEnabledCodeGeneration}.
However, we believe the full potential of reductions can only be exposed when
the effects are properly modeled on the dependence level.

The first to do so, namely to introduce reduction dependences, where Pugh and
Wonnacot~\cite{Pugh1994}. Similar to most other
approaches~\cite{Redon1993,Redon1994,Suganuma1996,Redon2000,Gautam2006,Sato2011,Zou2012},
the detection and modeling of the reduction was performed only on C-like
statements and utilizing a precise but costly access-wise dependence analysis
(see the upper part of Figure~\ref{fig:Evaluation}). In their work they utilize
both memory and value-based dependence information to identify statements with
an iteration space that can be executed in parallel, possibly after
transformations like array expansion. They start with the memory-based
dependences and compute the value-based dependences as well as the transitive
self-dependence relation for a statement in case the statement might not be
inherently sequential.

Stock et al.\cite{Stock2014} describe how reduction properties can be exploited
in the polyhedral model, however neither do they describe the detection nor how
omitting reduction dependences may affect other statements.

In the works of Redon and Feautrier~\cite{Redon1994} as well as the extension to
that by Gupta et al.~\cite{Gupta2002} the reduction modeling is performed on
SAREs on which array expansion~\cite{Feautrier88} and renaming was applied, thus
all dependences caused by memory reuse were eliminated. In this setting the
possible interference between reduction computation and other statements is
simplified but it might not be practical for general purpose compilers due to
memory constraints. As an extension to these scheduling approaches on SAREs we
introduced privatization dependences. They model the dependences between a
reduction and the surrounding statements without the need for any special
preprocessing of the input. However, we still allow polyhedral optimizations
that will not only affect the reduction statement but all statements in a
\pscop.

\subsection{Optimization} Optimization in the context of reductions is twofold.
There is the parallelization of the reduction as it is given in the input and
the transformation as well as possible parallelization of the input with
awareness of the reduction properties. The first idea is very similar to the
reduction-enabled code generation as described in
Section~\ref{sec:ReductionEnabledCodeGeneration}. In different variations,
innermost loops~\cite{Sato2011}, loops containing only a
reduction~\cite{Fisher1994,Pottenger1995} or recursive functions computing a
reduction~\cite{Xu04} were parallelized or replaced by a call to a possibly
parallel reduction implementation~\cite{Venkat2014}. The major drawback of such
optimizations is that reductions have to be computed either in isolation or with
the statements that are part of the source loop that is parallelized. Thus, the
reduction statement instances are never reordered or interleaved with other
statement instances, even if it would be beneficial. In order to allow
powerful transformations in the context of reductions, their effect, hence the
reduction dependences, as well as their possible interactions with all other
statement instances must be known.  The first polyhedral scheduling approach by
Redon and Feautrier~\cite{Redon1994} that optimally\footnote{e.g., according to
the latency} schedules reduction together with other statements assumed
reductions to be computable in one time step. With such atomic reduction
computations there are no reduction statement instances that could be reordered
or interleaved with other statement instances. Gupta et al.~\cite{Gupta2002}
extended that work and lifted the restriction on an atomic reduction
computation. As they schedule the instances of the reduction computation
together with the instances of all other statements their work can be seen as a
reduction-enabled scheduler that optimally minimizes the latency of the input.

To speed up parallel execution of reductions the runtime overhead needs to be
minimized. Pottenger~\cite{Pottenger1995} proposed to privatize the reduction
locations instead of locking them for each access and
Suganuma et al.~\cite{Suganuma1996} described how multiple reductions on the same
memory location can be coalesced. If dynamic reduction
detection~\cite{Rauchwerger1995} was performed, different privatization schemes
to minimize the memory and runtime overhead were proposed by Yu et
al.~\cite{Yu2004}. While the latter is out of scope for a static polyhedral
optimizer, the former might be worth investigating once our approach is extended
to multiple reductions on the same location.

In contrast to polyhedral optimization or parallelization, Gautam and
Rajopadhye~\cite{Gautam2006} exploited reduction properties in the polyhedral
model to decrease the complexity of a computation in the spirit of dynamic
programming. Their work on reusing shared intermediate results of reduction
computations is completely orthogonal to ours.

While Array Expansion, as introduced by Feautrier~\cite{Feautrier88}, is not a
reduction optimization, it is still similar to the privatization step of any
reduction handling approach.  However, the number of privatization copies the
approach introduces, the accumulation of these private copies as well as the
kind of dependences that are removed differ. While privatization only introduces
a new location for each processor or vector lane, general array expansion
introduces a new location for each instance of the statement.  In terms of
dependences, array expansion aims to remove false output and anti dependences
that are introduced by the reuse of memory while reduction handling approaches
break output and flow dependences that are caused by a reduction computation.
Because of the flow dependences---the actual reuse of formerly computed
values---the reduction handling approaches also need to implement a more
elaborate accumulation scheme that combines all private copies again.

\section{Conclusions and Future Work} \label{sec:ConclusionAndFutureWork}

Earlier work already utilized reduction dependences in different varieties,
depending on how powerful the detection was.  Whenever reductions have been
parallelized the reduction dependences have been implicitly ignored, in at least
two cases they have even been made explicit~\cite{Pugh1994,Stock2014}. However,
to our knowledge, we are the first to add the concept of privatization
dependences in this context. The reason is simple: we believe the parallel
execution of a loop containing a reduction is not always the best possible
optimization.  Instead we want to allow any transformation possible to our
scheduler with only one restriction: the integrity of the reduction computation
needs to stay intact. In other words, no access to the reduction location is
scheduled between the first and last instance of the reduction statement.  This
allows our scheduler not only to optimize the reduction statement in isolation,
but also to consider other statements at the same time without the need for any
preprocessing to get a SARE-like input.

To this end we presented a powerful reduction detection based on computation
properties and the polyhedral dependence analysis. Our design leverages the
power of polyhedral loop transformations and exposes various optimization
possibilities including parallelism in the presence of reduction dependences. We
showed how to model and leverage associativity and commutativity to relax the
causality constraints and proposed three approaches to make polyhedral loop
optimization reduction-aware.  We believe our framework is the first step to
handle various well-known idioms, e.g., privatization or recurrences, not yet
exploited in most practical polyhedral optimizers.

Furthermore, we showed that problems and opportunities arising from reduction
parallelism (see Section~\ref{sec:bicg_case_study}) have to be incorporated into
the scheduling process, thus the scheduling in the polyhedral model needs to be
done in a more realistic way. The overhead of privatization and the actual gain
of parallelism are severely influenced by the execution environment (e.g.,
available resources, number of processors and cores, cache hierarchy), however
these hardware specific parameters are often not considered in a realistic way
during the scheduling process.

Extensions to this work include a working reduction-aware scheduler and
the modeling of multiple reduction-like computations as well as other
parallelization preventing idioms. In addition we believe
that a survey about the applicability of different reduction detection schemes
as well as optimization approaches in a realistic environment is needed. In any
case this would help us to understand reductions not only from the
theoretical point of view but also from a practical one.

\section{Acknowledgments}
We would like to thank Tobias Grosser, Sebastian Pop and Sven Verdoolaege for
the helpful discussions during the development and implementation of this
approach. Furthermore, we want to thank the reviewers who not only provided
extensive comments on how to further improve this work but also pointed
us to related work that was previously unknown to us. Lastly, we would like to
thank Tomofumi Yuki for giving us many helpful tips.

 \BIBLIOGRAPHYSTYLE
\BIBLIOGRAPHYFILE

\end{document}